# Surface and bulk components of electrical conductivity in the (presumably special topological) Kondo insulator SmB$_6$ at lowest temperatures

S. Gabáni[a], G. Pristáš[a], I. Takáčová[a], N. Sluchanko[b], K. Siemensmeyer[c], N. Shitsevalova[d], V. Filipov[d] and K. Flachbart[a*]

[a] Institute of Experimental Physics, Slovak Acad. Sciences, SK-04001 Košice, Slovakia

[b] General Physics Institute of RAS, RU-119991 Moscow, Russia

[c] Helmholtz Zentrum Berlin, D-14109 Berlin, Germany

[d] Institute for Problems of Materials Science of NASU, UA-03680 Kiev, Ukraine

**Abstract**

Samarium hexaboride (SmB$_6$) has recently been considered to be a topological Kondo insulator (TKI), the first strongly correlated electron system to exhibit topological surface conduction states. In this contribution, results of electrical resistivity measurements $\rho(T)$ between 80 K and 0.08 K of various SmB$_6$ single crystalline samples are presented, analyzed and discussed. The received results imply that the residual conductivity of SmB$_6$ below about 4 K is of non-activated (metallic-like) nature. It is shown that this metallic-like behaviour can be attributed both to surface (2D) conduction states, as may be expected in case of a topological insulator, as well as to the highly correlated many-body (3D) bulk ground state which is formed within the gap of this compound. From this it follows that in SmB$_6$, where surface conductivity states are clearly present, there is in parallel also a bulk contribution to residual electrical conductivity originating from the strongly correlated electron system with valence fluctuations. This raises the question whether SmB$_6$ does not form a new / special type of topological insulator in which in the energy gap besides the surface conduction states, there is also a conducting narrow in-gap band originating from the bulk strongly correlated electron system.

## 1. Introduction

Samarium hexaboride SmB$_6$ has been since many years considered to be a typical representative of intermediate valence semiconductors, in which the oxidation state of samarium ions fluctuates between the Sm$^{2+}$ (4f$^6$) and Sm$^{3+}$ (4f$^5$5d$^1$) configurations, and as a prime example of a narrow-gap Kondo insulator [1-3]. More than four decades after the pioneering investigations of Nickerson et al. [4] a large number of detailed and rather comprehensive studies of SmB$_6$ have been performed. However, some fundamental properties of SmB$_6$ such as the electrical transport mechanism (electrical conductivity) at very low temperatures are still not fully understood. Experimental investigations on SmB$_6$ (see e.g.

---

[*] Corresponding author: flachb@saske.sk

refs. 5-12) have shown that in the energy spectrum of this material at least three energy scales and regimes of low temperature electron kinetics exist. In the temperature range 70 K < $T$ < 15 K the properties of $SmB_6$ are governed by the hybridization gap $E_g \approx$ 10 - 20 meV which results from hybridization between itinerant conduction electrons and localized $f$ electrons. At lower temperatures, between about 15 and 4 K, a narrow in-gap band separated from the bottom of the conduction band by an activation energy of $E_d \approx$ 3 - 5 meV has been observed. The properties of this narrow band seem to be influenced by the content of impurities and imperfections of the specific sample. Below about 4 K the electrical conductivity saturates, indicating a residual conductivity / resistivity which remains finite as $T \rightarrow$ 0 K, and implies that the Fermi level $E_F$ lies / is pinned within the $E_d$ in-gap states.

Various models have been proposed to explain the formation of the $E_d$ band and the origin of the residual conductivity, but so far no final conclusion could be obtained. Assuming that the transport is dominated by the three-dimensional (3D) bulk, the residual conductivity can be attributed to electron transport in a correlated many-body ground state formed within the $E_d$ intra-gap band [10,13,14] of this strongly correlated electron compound.

Another approach which could solve the problem of high residual resistivity in $SmB_6$ may lie in a recent theoretical prediction [15-18] that Kondo insulators can be topological insulators. Topological insulators, which have been the subject of intense theoretical and experimental investigation over the last years, are insulating in the bulk but have in their gap metallic surface states. These states are robust, being protected by time-reversal symmetry. The TKI approach for $SmB_6$ has been recently supported by a number of various experimental studies (see e.g. [19-21]). Specifically in ref. 20, where transport properties were investigated on a sample with specialized geometry and eight coplanar electrical contacts (four on each side of the sample), below 4 K a crossover from bulk to surface conduction with a fully insulating bulk was observed.

In this paper we present results of detailed electrical resistivity measurements performed on five various $SmB_6$ single crystals with the aim to contribute to determination of the origin of charge transport, particularly in the lowest temperature range. Results received on some of these samples have been already published before (see e.g. ref. 22), however, in this paper they are used to compare them with the latest results and to undertake the corresponding analysis.

## 2. Experimental

The resistivity measurements between 80 mK and 2 K were carried out in a dilution refrigerator, and between 1.6 and 80 K in a $^4$He cryostat. In the dilution refrigerator the resistance of the samples was measured by an ac-resistance bridge, in the $^4$He cryostat it was determined by an ac-current source and a lock-in nanovoltmeter.

In both cases, a standard four-terminal method applied from both sides of the sample was used (as in ref. 20) as a conventional four-terminal lateral measurement $R_{lat}$ using contacts only from one side cannot distinguish whether the conduction at low temperatures is bulk- or surface-dominated (see Fig. 1). Specifically, we made therefore also vertical measurements $R_{vert}$ by passing current from one front-side contact to the back-side contact directly opposite, and measuring the voltage using a different set of opposing front-side and back-side contacts. And, we also made a hybrid measurement $R_{hybr}$ by passing current through two front-side contacts as in the lateral measurement, but measuring the voltage on two back-side contacts. If in these cases the resistivity plateau is a bulk transport phenomenon, the resistance should be proportional to the resistivity for all three measurement configurations, each with a different proportionality constant. However, if the plateau appears due to surface conduction, these three four-terminal resistances should behave differently as a function of temperature.

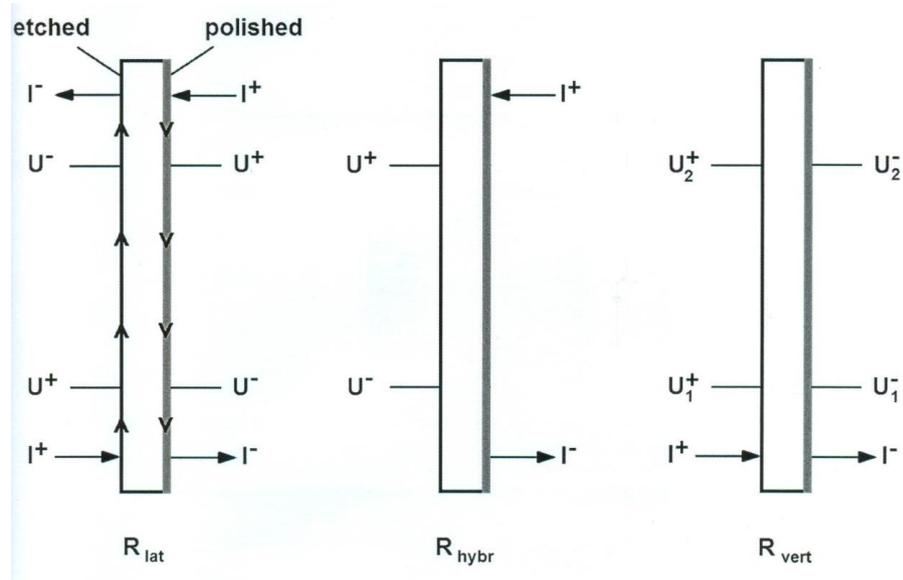

Fig. 1: Cross section of the investigated sample with applied electrical contacts.

The single crystalline $SmB_6$ samples were prepared from starting materials of different purities using the floating zone melting technique. The dimension of the sample used for $R_{lat}$, $R_{vert}$, and $R_{hybr}$ measurements was about 0.2x1.4x5 mm$^3$. Both current and voltage contacts were made using silver paste. As thermometers, commercially calibrated *Lake Shore* sensors were used.

## 3. Results and discussion

In Fig. 2 resistances denoted above as $R_{lat}$, $R_{hybr}$ and $R_{vert}$ of the sample (sample 5 - to which contacts from both sides were applied) are shown as a function of temperature. Similarly as in ref. 20, at lowest temperatures the values of $R_{hybr}$ (in this configuration the voltage contacts of the opposite surface are used) are lower than the corresponding $R_{lat}$ (standard method) values, and the values of $R_{vert}$ higher than these of $R_{lat}$. The discrepancies / disagreements between those results (see Fig. 2) and the results of [20] may follow from slightly different sample and contact geometries used in specific measurements. These results thus demonstrate the important role of the surface electrical conduction at lowest temperatures and support the conclusions of various investigations which show that $SmB_6$ exhibits features a topological insulator.

Let us now analyze and discuss further results. The temperature dependence of the electrical resistivity $\rho(T)$ (based on $R_{lat}$ measurements) of all investigated $SmB_6$ samples (samples 1, 2, 3 and 4 were measured already earlier) up to 80 K is shown in Fig. 3.

In case of all five samples, before their insertion into the cryostat, they were etched in dilute nitric acid, which lead to their mat (not shiny) surfaces. Shown by broken line in Fig. 3 is the dependence of sample 2 after polishing (having a shiny / mirror like surface at room temperature), which reduced the residual resistivity of this sample twice. In case of sample 5, polishing of sample surface lead to a reduction of its residual resistivity (or to an increase of its residual conductivity) also by a factor of about 2.

The observed dependencies (Fig. 3) of all samples show an approximately exponential increase of resistivity as the temperature is lowered from 80 K to 5 K indicating a gap structure in the density of states. Below about 4 K a saturation of resistivity (usually with a very slight increase of $\rho(T)$ towards the lowest temperatures) is observed. To derive more information about the gap and about the carrier kinetics at various temperatures, the local activation energy as a function of temperature was determined from the above data.

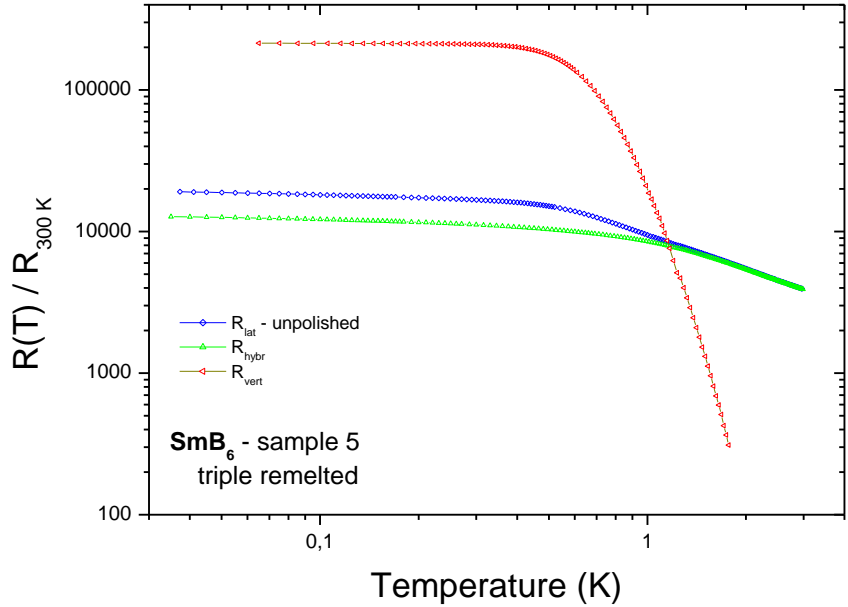

Fig. 2: Temperature dependences of sample 5 resistances denoted as $R_{lat}$, $R_{hybr}$ and $R_{vert}$.

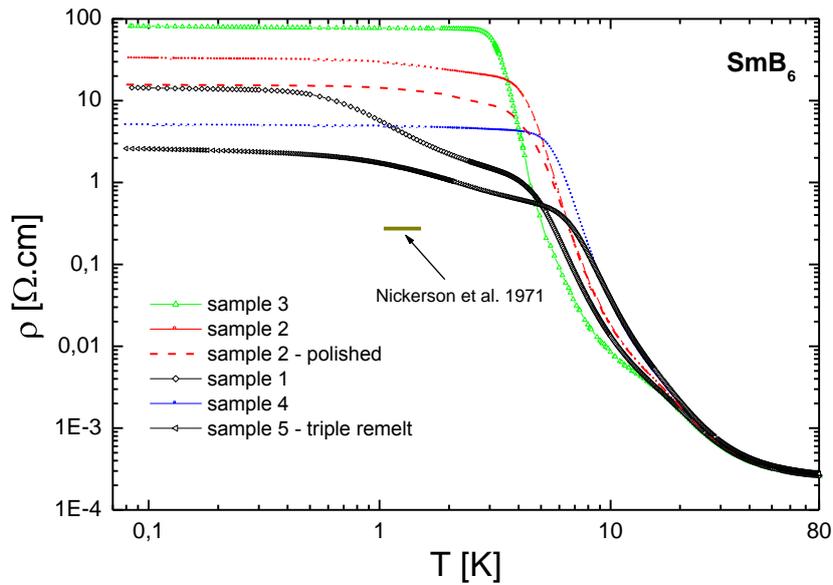

Fig. 3: Temperature dependence of the electrical resistivity of investigated SmB$_6$ samples. The broken red line shows the resistivity of sample 2 after polishing.

The local activation energy defined as $W(T) = d(\ln\rho(T))/d(1/k_B T)$ was computed by means of a standard numerical derivative analysis and is plotted in Fig. 4. As can be seen from this figure, at temperatures around 20 - 30 K the conductivity is characterized by an activation energy with a value of about 5.6 meV. This value can be attributed to semiconductor intrinsic conductivity of the form $\rho(T) \propto \exp(E_g/2k_B T)$, where $E_g \approx 11.2$ meV is in case of $SmB_6$ the width of the hybridization gap. Fig. 4 shows that $E_g$ has the same value and is seen in the same temperature range in all samples. The $W(T)$ lowering observed above 30 K probably corresponds also with the strong decrease of Hall mobility in this temperature range [9].

Between 10 K and 3 K a second pronounced activation energy appears. Its value $E_d$ varies between ~3.7 and ~5.6 meV, and seems to depend on the quality of the sample (concentration of impurities in the sample, stoichiometry of the sample), and the direction of measuring current [23]. The same applies for the temperature at which this activation energy becomes fully developed (see the maxima of the $W(T)$ dependences) - it appears at lower temperature for samples with a higher residual resistivity.

With further decrease of temperature the calculated activation energy decreases to very low values, and below about 1 K, $W$ is already much lower than the available thermal energy $k_B T$. The residual conductivity at lowest temperatures therefore shows to non-activated (metallic / metallic-like) transport within the in-gap states.

This decrease / drop of the activation energy, which is most rapid for the purest sample (where a decrease of $W$ by about 4 orders of magnitude is observed) can be (from the 3D-bulk point of view) considered as "metallization" or as formation of a heavy fermion / strongly correlated electron system of delocalized carriers within the in-gap states of the $E_d$ band. The Fermi level $E_F$ has then to be pinned in this band.

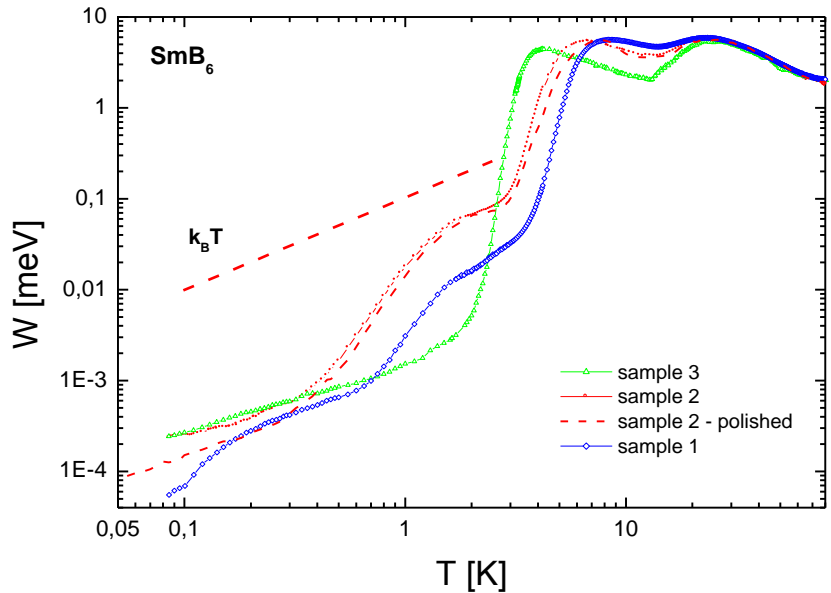

Fig. 4: The temperature dependence of the local activation energy $W(T)$ of $SmB_6$ samples calculated from resistivity data displayed in Fig. 1.

The observed anomalies at about 1.5 K for samples 1 and 2 originate probably from an additional narrow band which is formed in the energy gap (close to the $E_d$ - band) due to a larger concentration of impurities and imperfections in the sample. Such a band has not observed in the purest sample (sample 3).

Surface states, in which the samarium ions (being in non-stoichiometric positions) are expected to be in non-intermediate (stable) valence $Sm^{3+}$ - states [1], lead to an increase of the concentration of conduction electrons at the surface, and thus to emergence of surface conductivity. The reason why at the surface $Sm^{+3}$- states instead of $Sm^{+2}$- states are stabilized may lie in the fact that electrons at the surface have a lower zero-point energy when they are spread out through the (large) surface (the case with $Sm^{+3}$ ions) than when they are confined to one atom (in the case with $Sm^{+2}$ ions).

The influence of different surface states to residual conductivity was (as mentioned above) in a simple way investigated on samples 2 and 5 which were measured both with their surface being etched (which lead to a mat / not shiny surface) or polished by a fine sandpaper (which lead to a shiny / mirror-like surface). The results for sample 2 displayed in Figs. 3 and 4 show that polishing leads to a higher residual conductivity (by about a factor of 2), to a slight shift of the "metallization" process to higher temperatures, and to a smaller activation energy $W$ in the lowest temperature range. However, the shapes (character) of $\rho(T)$ and $W(T)$ dependencies do not change essentially. Same results were received also on sample 5.

As follows from above presented results, from bulk measurements on good quality single crystalline samples of $SmB_6$ above 4 K one can deduce the presence both of the indirect gap ($E_g$) and of the activation energy of intra-gap states ($E_d$). Therefore, these bulk characteristics can be considered as robust in $SmB_6$. It is worth to note that similar $E_g$ and $E_d$ values were deduced also for $SmB_6$ also by point-contact spectroscopy studies (see e.g. [24]) which is also a surface sensitive technique. Thus, $E_g$ and $E_d$ as bulk characteristics are robust in $SmB_6$.

Moreover, within almost constant values of $E_g$ and $E_d$, provides the variation of the very low temperature residual electrical conductivity / resistivity in $SmB_6$ values between about 0.3 Ohm.cm at 1 K (see e.g. Nickerson et al. 1971 [4]) and about 80 Ohm.cm (see e.g. sample 3 in Fig. 3) a rather large change - of more than 250 - times. On the other hand, changing the $SmB_6$ surface by etching or polishing, i.e. between mat / not shiny and shiny / mirror like, respectively, leads to a change of residual resistivity / conductivity by only about 2 - times. Thus, the rather large variation of residual resistivity at lowest temperatures (of more than 250 - times) most probably cannot be attributed to different surface states, but to a bulk electron effect. In relation with this the exciton condensation proposed in [25,26] looks to be the natural mechanism, taking into account that fast on-site 4f-5d charge/valence fluctuations (the average valence of Sm-ions in $SmB_6$ is about $Sm^{+2.6}$) are followed by a formation of short range (~5A) exciton-polaron many-body states in the temperature range below about 15 K, where the in-gap states ($E_d$) are formed and within which the residual electrical conductivity originates. Experimentally the plausibility of this model was investigated and supported in [23]. These findings point to an important role of the bulk nature of residual electrical conductivity in $SmB_6$. However, in parallel also the $Sm^{+3}$ produced surface conduction states make a contribution to the residual conductivity.

In Fig. 5 the change of the $SmB_6$ resistance at 0.2 K in perpendicularly applied magnetic field is shown. As in similar measurements at liquid helium and intermediate temperatures (see e.g. [7]), a negative magnetoresistance was observed (in [27] it was, moreover, determined that the insulating gap of $SmB_6$ closes at very high magnetic field of about 80 - 90 T). The observed negative magnetoresistance seems to contradict the TKI model for $SmB_6$ as in a 2D surface (almost) free electron system due to the Lorentz force an increase of resistance should be expected (as e.g. in other / typical topological insulators like e.g. $Bi_2Se_3$, $Bi_2Te_3$ and $Bi_2(TeSe)_3$).

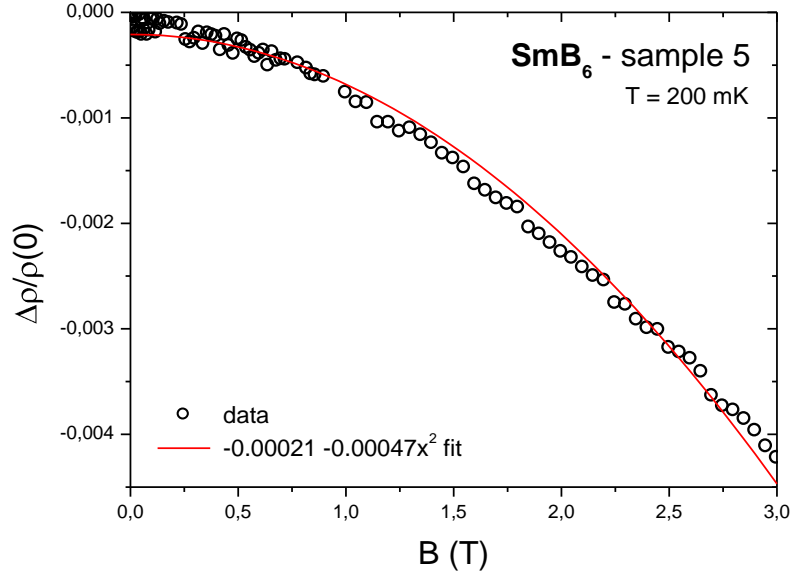

Fig. 5: Magnetoresistance of sample 5 at 0.2 K in perpendicularly applied magnetic field.

Taking into account also some other previous transport measurements on $SmB_6$, in the region of residual resistivity (below about 4 K), both the Seebeck coefficient and the Hall mobility of charge carriers there decrease dramatically and become practically equal to zero [28]. These results are most probably not in accordance with predictions of the TKI model applied to $SmB_6$.

Moreover, a detailed study of the $SmB_6$ low energy electrodynamics in the spectral range 0.6 - 4.5 meV performed in [10] gives also evidence for a narrow in-gap band with activation energy of ~3 meV. And it should be stressed, that these measurements in the far-infrared spectral range at T = 3 K detected only bulk (not surface) electrical conductivity and dielectric permittivity of this compound. Therefore, also from these measurements the in-gap states in $SmB_6$ seem to manifest themselves above all as bulk states.

## Conclusions

In summary, the received results imply that the residual conductivity / resistivity of $SmB_6$ below about 4 K is of non-activated (metallic-like) nature. This metallic-like behaviour cannot be most probably attributed only to features of charge carriers transport through the topological surface states in $SmB_6$. On the contrary, two conduction channels seem to provide the saturated low temperature electrical conductivity: the charge transport through the surface (2D) conduction states coming from $Sm^{+3}$ ions, as may be expected in case of a topological insulator, as well as the charge transport through highly correlated many-body bulk (3D) ground state formed within the intra-gap states of this compound, which originate from valence fluctuations of Sm-ions, i.e. from the competition between electron localization and delocalization in this strongly correlated electron system. From these results thus follows, that in $SmB_6$ both these components of conductivity are important.

The received results also raise the question whether $SmB_6$ is not a special type of topological insulator in which in the energy gap, besides the 2D surface conduction states (with a roughly parabolic energy vs. momentum dependence) there is concurrently also a conducting narrow in-gap band (with roughly constant energy) originating from the 3D bulk strongly correlated electron system.

## Acknowledgements

This work was supported by projects VEGA 2/0106/13, APVV 0132-11, and by CFNT MVEP project of the Slovak Academy of Sciences.